\documentclass[fleqn,11pt,letterpaper]{article}

\usepackage{amsfonts}
\usepackage{amssymb}
\usepackage{amsmath}
\usepackage{amsthm}
\usepackage{enumitem}
\usepackage{multirow}
\usepackage{float}
\usepackage{graphicx}
\usepackage{epstopdf}
\usepackage{color}
\usepackage[margin=2.57cm]{geometry}
\definecolor{darkred}{rgb}{0.5,0.2,0.2}
\usepackage[pdftitle={Monitoring the Global Carbon Budget},pdfauthor={Mikkel Bennedsen},colorlinks=TRUE,allcolors=darkred]{hyperref}
\usepackage{booktabs}
\usepackage{pdflscape}
\usepackage[round]{natbib}

\usepackage{tabularx}

\usepackage{float}
\floatstyle{plaintop}
\restylefloat{table}

\restylefloat{table}

\theoremstyle{plain}

\theoremstyle{definition}

\theoremstyle{remark}

\newcommand{\R}{\mathbb{R}}

\def\bi{\begin{itemize}}
\def\ei{\end{itemize}}

\allowdisplaybreaks
\graphicspath{{Figures/}}
\numberwithin{equation}{section}

\newif\ifi

\usepackage{lineno}

\begin{document}

\title{Designing a statistical  procedure for monitoring global carbon dioxide emissions}

\author{Mikkel Bennedsen\thanks{
Department of Economics and Business Economics and CREATES, 
Aarhus University, 
Fuglesangs All\'e 4,
8210 Aarhus V, Denmark.
E-mail:\
\href{mailto:mbennedsen@econ.au.dk}{\nolinkurl{mbennedsen@econ.au.dk}}. 
}}

\maketitle

\begin{abstract}
Following the Paris Agreement of $2015$, most countries have agreed to reduce their carbon dioxide (CO$_2$) emissions according to individually set Nationally Determined Contributions. However, national CO$_2$ emissions are reported by individual countries and cannot be directly measured or verified by third parties. Inherent weaknesses in the reporting methodology may misrepresent, typically an under-reporting of, the total national emissions. This paper applies the theory of sequential testing to design a statistical monitoring procedure that can be used to detect systematic under-reportings of CO$_2$ emissions. Using simulations, we investigate how the proposed sequential testing procedure can be expected to work in practice. We find that, if emissions are reported faithfully, the test is correctly sized, while, if emissions are under-reported, detection time can be sufficiently fast to help inform the $5$ yearly global ``stocktake'' of the Paris Agreement. We recommend the monitoring procedure be applied going forward as part of a larger portfolio of methods designed to verify future global CO$_2$ emissions.
\end{abstract}

\vspace*{1.5em}


\vspace*{1.5em}



\newpage

\section{Introduction}
The Paris Agreement of $2015$ instituted a transnational commitment to limit global temperature rise to $2.0$ degrees centigrade, and preferably $1.5$ degrees centigrade, above pre-industrial levels  \citep{FCCC2015}. It is widely accepted that to achieve this goal, substantial reductions of anthropogenic carbon dioxide (CO$_2$) emissions are needed \citep{SONBT2016,Luderer2018,Millar2018,TG2018,TON2018}. Indeed, the recent report from the Intergovernmental Panel on Climate Change (IPCC)  states that to stay below $1.5^\circ$C, emissions should be reduced by almost half by $2030$ (from $2010$ levels) with a level close to zero in $2050$ \citep[][Chapter 2]{IPCC15}.

Reducing emissions substantially requires all nations to work towards this goal, particularly the nations that are currently emitting the most \citep{Meinshausen2015}. The Paris Agreement therefore requires signing parties to deliver mandatory annual emissions reports, which are to be assessed during $5$ yearly ``stocktakes'' of the global emissions status. Unfortunately, since data on CO$_2$ emissions are \emph{reported} by the nations themselves, instead of being \emph{measured} by the global community, this could create incentives for individual nations to under-report emissions \citep{Peters2017}. In this way, nations that are not living up to their Paris commitments could, by under-reporting their CO$_2$ emissions, nevertheless appear to be fulfilling their Nationally Determined Contribution targets. This is especially worrisome, as some countries have notoriously opaque emissions reporting and verification practices \citep{GLZH2012,DGPR2013,GlobalCorruption2013,GZ2014,KPA2016,Nature_Editorial2018,Zhang2019}. Indeed, the problem of verifying the reported CO$_2$ emissions was one of the key topics discussed at the recent Conference of the Parties meeting in Katowice, Poland \citep[COP24;][]{Katowice2018}.

 The aim of this paper is to design a statistical procedure that can help in verifying reported anthropogenic CO$_2$ emissions. To do this, we exploit the idea of a balanced carbon budget \citep{GCB2020short}: because the Earth's carbon cycle is a closed system, the amount of anthropogenically emitted CO$_2$ must equal the amount of CO$_2$ absorbed in the three carbon sinks, namely the atmosphere, the terrestrial biosphere, and the oceans. This insight gives rise to the carbon budget equation \citep{GCB2020short}
 \begin{linenomath}
\begin{align}\label{eq:be0}
E_t^\textrm{FF} + E_t^\textrm{LUC} = G_t^\textrm{ATM} + S_t^\textrm{OCN} + S_t^\textrm{LND} + B_t^\textrm{IM},
\end{align}
\end{linenomath}
where $E_t^\textrm{FF}$ and $E_t^\textrm{LUC}$ are year-$t$ CO$_2$ emissions from fossil fuel burning and land-use change, respectively, and $ G_t^\textrm{ATM},$ $S_t^\textrm{OCN}$, $S_t^\textrm{LND}$ denote the year-$t$ uptake of CO$_2$ in the atmosphere, the oceanic carbon sink, and the terrestrial (``land'') carbon sink, respectively. The quantity $B_t^\textrm{IM}$, dubbed the \emph{budget imbalance}, is a residual term introduced to balance Equation \eqref{eq:be0}, and thus captures deviations in the carbon budget equation due to mis-measurements of the remaining quantities in \eqref{eq:be0}. Such mis-measurements can be transitory, arising e.g. from measurement errors, or they can be systematic, arising either from systematic biases in the models used to estimate the carbon flux data series,  $E_t^\textrm{LUC}$, $S_t^\textrm{OCN}$, and $S_t^\textrm{LND}$, or, importantly for this paper, systematic mis-reportings of anthropogenic CO$_2$ emissions from fossil fuels, $E_t^\textrm{FF}$. (Since data on atmospheric CO$_2$ levels are from direct observations, systematic mis-measurement of $G_t^\textrm{ATM}$ is unlikely.)

The first contribution of this paper is an extensive statistical analysis of the time series of the budget imbalance $B_t^\textrm{IM}$, constructed using the data supplied by the Global Carbon Project (Section \ref{sec:data} contains details on the data). We find that these data are historically well-described by a zero-mean stationary process. Adjacent data points in the budget imbalance are positively correlated, suggesting that mis-measurements of the carbon flux data series tend to persist over time, a point also noted in \citet[][pp. 3295--3296]{GCB2020short}. However, the fact that the budget imbalance data appear to be stationary and have zero mean, provides evidence that the mis-measurements of the carbon fluxes in Equation \eqref{eq:be0}, although persistent, are ultimately of transitory nature. In other words, our analysis suggests that the historical climate flux data $E_t^\textrm{FF}, E_t^\textrm{LUC},  G_t^\textrm{ATM}, S_t^\textrm{OCN}$, and $S_t^\textrm{LND}$ have been compatible with the global carbon budget equation \eqref{eq:be0}. In particular, the analysis suggests that up until now, CO$_2$ emissions have been reported without systematic biases. To our knowledge, ours is the most thorough statistical analysis of the budget imbalance data to date and the findings might be of independent interest. We then show that if, sometime in the future, emissions are \emph{not} truthfully reported, the observed budget imbalance $B_t^\textrm{IM}$ will undergo a \emph{structural break}: some process with non-zero mean will be introduced into these data.

The second contribution of the paper is to use these properties of the budget imbalance data to develop a monitoring procedure, which can be used for detecting potential future under-reportings of CO$_2$ emissions. Our procedure relies on a test statistic, derived from the residuals of the global carbon budget, i.e. $B_t^\textrm{IM}$. In effect, we sequentially test the null hypothesis that the future climate flux data $E_t^\textrm{FF}, E_t^\textrm{LUC},  G_t^\textrm{ATM}, S_t^\textrm{OCN}$, and $S_t^\textrm{LND}$ are compatible with the global carbon budget equation \eqref{eq:be0}. If, for some future time period $t$, this null hypothesis can be rejected, we conclude that there is evidence of systematic mis-measurements of the variables in Equation \eqref{eq:be0}. In particular, a rejection of the null can provide evidence for $E_t^\textrm{FF}$ being systematically under-reported.

The theory of sequential testing goes back at least to \cite{page1954} and have since been successfully applied in many areas, such as economics \citep[][]{CSW1996}, finance \citep[][]{AHHHS2012}, engineering \citep[][]{Lai1995}, and medicine \citep[]{Unkel2012}. While non-sequential (``off-line'') testing methods have been widely applied for detecting structural breaks in climate-related data \citep[see][for a review]{Reeves2007}, sequential methods do not appear to have been widely considered in the climate literature. An exception is \cite{HHKS2004}, where sequential testing methods are applied to detect structural breaks in temperature data. Existing sequential testing procedures generally rely heavily on the statistical properties of the data, usually the so-called long run variance. However, the nature of the available climate data means that, in the context of this paper, the long run variance is not well-defined.  The reason for this is that, in addition to a new data point being added to our data set each year, previous data points might be revised, i.e. retroactively updated (see Section \ref{sec:data} and especially Figure \ref{fig:gcb} below), which means that the statistical properties of the key object of this paper, the budget imbalance, might change from year to year. This leads us to design a ``pivotal'' sequential test statistic, which, asymptotically, does not rely on the statistical properties of the data and which can thus be used in the setup studied here.

In a simulation experiment, we illustrate the use of the sequential testing procedure proposed in the paper and investigate its finite sample performance. We find that, under realistic conditions, the test is correctly sized under the null, i.e. when CO$_2$ emissions are reported without systematic bias. The empirical (power) properties of the test when the alternative is true, i.e. when CO$_2$ emissions are under-reported, depend on the magnitude of under-reporting. Our simulations show that when the magnitude of under-reporting is small, misreporting can be difficult to detect in practice. This is in line with the conclusions reached in a related, but less statistically rigorous, approach considered in \cite{Peters2017}. For moderate-to-large magnitudes of under-reporting, however, the power of the test (probability of correctly detecting misreporting)  is close to one and mean detection time of the method is on the order of $5$--$10$ years. Consequently, the method proposed in this paper can potentially help the global community in future efforts of verifying reported CO$_2$ emissions.

The rest of the paper is structured as follows. Section \ref{sec:gcb} presents the global carbon budget equation and introduces the data we use. The section also contains statistical analyses of the historical budget imbalance (Section \ref{sec:bi}) and explores how the budget imbalance might evolve in the future, should emissions become under-reported (Section \ref{sec:mis}). Section \ref{sec:design} presents the proposed monitoring procedure and the details on how it can be implemented in practice. The section also summarizes the results of a number of simulation exercises (Section \ref{sec:sims}). Lastly, Section \ref{sec:summary}  discusses the findings of the paper and provides an outlook on how the methods of this paper can be used in practice going forward. An Electronic Supplementary Material file is available online.\footnote{\url{https://sites.google.com/site/mbennedsen/research/monitoring}.}

\section{The global carbon budget}\label{sec:gcb}
As mentioned in the introduction, the closedness of the Earth's carbon system implies the carbon budget equation \eqref{eq:be0}, where $E_t^\textrm{FF}$ is CO$_2$ emissions from fossil fuel burning, cement production, and gas flaring; $E_t^\textrm{LUC}$ is CO$_2$ emissions from land-use change (deforestation); $G_t^\textrm{ATM}$ is growth of atmospheric CO$_2$ concentration; $S^\textrm{OCN}_t$ is the flux of CO$_2$ from the atmosphere to the oceans; and $S^\textrm{LND}_t$ is the flux of CO$_2$ from the atmosphere to the terrestrial biosphere. 

Essentially, the carbon budget equation \eqref{eq:be0} is an accounting identity: it states that anthropogenic CO$_2$ emissions (left hand side) must equal the fluxes of CO$_2$ into the three carbon sinks (right hand side), namely the atmosphere ($G_t^\textrm{ATM}$), the oceans ($S_t^\textrm{OCN}$), and the terrestrial biosphere ($S_t^\textrm{LND}$). The \emph{budget imbalance}
\begin{linenomath}
\begin{align}\label{eq:bi}
 B_t^\textrm{IM} = E_t^\textrm{FF} + E_t^\textrm{LUC} - G_t^\textrm{ATM} - S_t^\textrm{OCN} - S_t^\textrm{LND}
\end{align}
\end{linenomath}
is implicitly defined such that the carbon budget equation \eqref{eq:be0} is balanced when inserting data for the carbon flux variables $E_t^\textrm{FF}, E_t^\textrm{LUC}, G_t^\textrm{ATM}, S_t^\textrm{OCN}$, and $S_t^\textrm{LND}$.  Indeed, in theory, i.e. if we could get completely accurate measurements of these carbon fluxes, the carbon budget would be balanced, so that $B_t^\textrm{IM}=0$ for all $t$. However, due to measurement errors in the data, the budget imbalance will in general be non-zero, i.e. $B_t^\textrm{IM} \neq 0$. Section \ref{sec:bi} contains statistical analyses of data on the budget imbalance, showing that, historically, it is well-represented by a mean-zero stationary process; Section \ref{sec:mis} shows that in the case of systematic mis-measurements of the quantities in \eqref{eq:be0}, such as what might arise if CO$_2$ emissions become misreported, this will cease to be the case. First, the following section presents some details regarding the data we use for the carbon flux variables $E_t^\textrm{FF}, E_t^\textrm{LUC}, G_t^\textrm{ATM}, S_t^\textrm{OCN}$, and $S_t^\textrm{LND}$.

\subsection{Data}\label{sec:data}
Each year, The Global Carbon Project\footnote{\url{http://www.globalcarbonproject.org/}} (GCP) publishes a report, the ``Global Carbon Budget'', on the state of the global carbon cycle, taking its departure point in the global carbon budget and its equation \eqref{eq:be0}. With each report, an up-to-date data set of the carbon flux variables in \eqref{eq:be0} is supplied. It is these data that we study in this paper. As of this writing (in February $2021$), the GCP has released four of these annual reports with data on each of the quantities in \eqref{eq:be0}, namely on the Global Carbon Budget in the years 2017, 2018, 2019, and 2020 \citep[see][respectively]{GCB2017short,GCB2018short,GCB2019short,GCB2020short}.

The fossil fuel emissions data $E_t^\textrm{FF}$ are compiled from \cite{gilfillan2019} and \cite{UNFCCC}. These data are constructed from national replies to the Annual Questionnaire on Energy Statistics \citep{UNstats} conducted by the U.N. Statistics Division and released in the annual Energy Statistics Yearbook \citep{ESyearbook}. Hence, the raw data consists of reported energy statistics, which are then converted into estimates of CO$_2$ emissions using the approach developed in \cite{MR1984}. The land-use change emissions data $E_t^\textrm{LUC}$  are obtained as the average of separate estimates coming from global climate models \citep{HDP15,HN17,LUC2020}. The growth rate in atmospheric CO$_2$ data $G_t^\textrm{ATM}$ are based on observations (measurements) of the concentration of CO$_2$ in the atmosphere at several locations on the globe \citep{DT18}. The sink data $S_t^\textrm{OCN}$ and $S_t^\textrm{LND}$ are model-based, i.e. data on these series are constructed from the output of various climate models. In the latest version of the GCP report, GCB2020, the data on the ocean sink flux $S_t^\textrm{OCN}$ are averages of the output from $9$ different models and the land sink data $S_t^\textrm{LND}$ are averages of the output from $17$ different models. We refer to  \citet[][pp. 3274--3288]{GCB2020short} and the papers therein for more in-depth explanation on the construction of the individual data series.

Each year, when the GCP publishes their report, the data set is updated with the newest available data so that each report adds a (yearly) data point for each quantity in \eqref{eq:be0}. However, the nature of the data, as discussed in the previous paragraph, means that the data set might also be subject to \emph{revisions}. That is, in a given year, the GCP data set will not only contain a new data point, but it might also update old data points. Such updates, or revisions, might be due to a number of factors. For instance, revisions in $E_t^\textrm{FF}$ might result from revisions of fossil fuel consumption from individual countries (e.g. from changes in accounting procedures) or from the method of converting fossil fuel consumption into CO$_2$ emissions. The data series for $E_t^\textrm{LUC}$, $S_t^\textrm{OCN}$, and $S_t^\textrm{LND}$ are averages of model-based estimates and revisions in these data might result from changes in the underlying models and/or additions or removals of models used to construct the averages. For instance, new insights into the underlying drivers of the sinks might improve the model-based estimates of $S_t^\textrm{OCN}$ and $S_t^\textrm{LND}$ in the future. Similarly, since the data on atmospheric CO$_2$ emissions growth, $G_t^\textrm{ATM}$, are constructed from measurements of the levels of CO$_2$ in the atmosphere at several locations on the planet, these data might be revised if the method for constructing $G_t^\textrm{ATM}$ changes.

\begin{figure}[t!]
\centering
\includegraphics[width=1.0\linewidth]{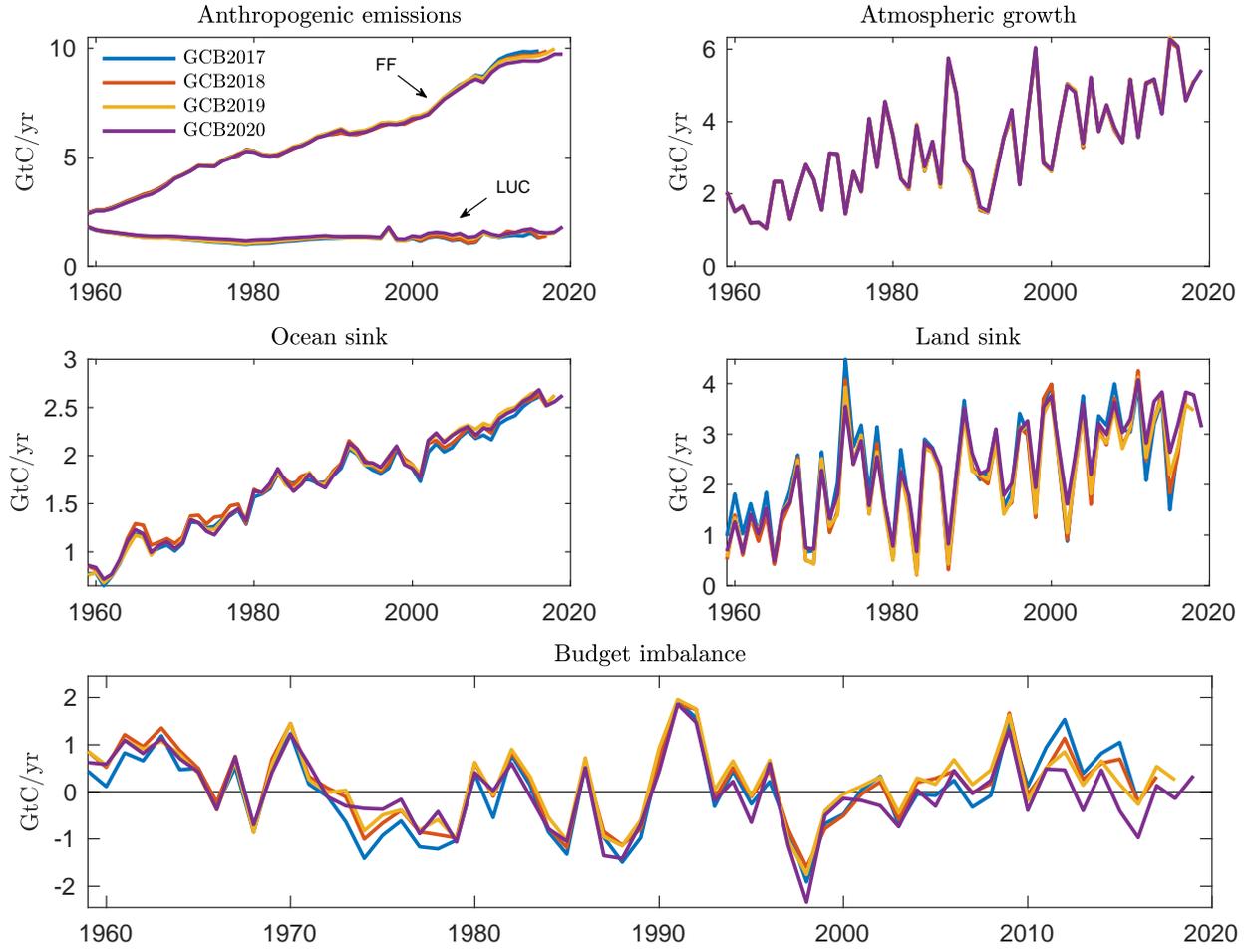} 
\caption{\emph{Time series data of the carbon fluxes of the carbon budget \eqref{eq:be0} from three vintages of the GCP data set. Top left: Anthropogenic emissions, $E_t^\textrm{FF}	$ and $E_t^\textrm{LUC}$. Top right: Atmospheric growth, $G_t^\textrm{ATM}$. Mid left: Ocean sink flux, $S_t^\textrm{OCN}$. Mid right: Terrestrial sink flux, $S_t^\textrm{LND}$. Bottom: Budget imbalance, $B_t^\textrm{IM}$. }}
\label{fig:gcb}
\end{figure}

The top two rows of Figure \ref{fig:gcb} plot the data series $E_t^\textrm{FF}$, $E_t^\textrm{LUC}$, $G_t^\textrm{ATM}$, $S_t^\textrm{OCN}$, and $S_t^\textrm{LND}$ from the four GCP data series ``GCB2017'', ``GCB2018'', ``GCB2019'', and ``GCB2020'', published along the four reports \cite{GCB2017short}, \cite{GCB2018short}, \cite{GCB2019short}, and \cite{GCB2020short}.\footnote{\label{foot:carbonation}In the $2020$ version of the reports, GCB2020, a new sink term, $S_t^C$, was introduced into the budget equation, which is an estimate of the carbon sink from cement carbonation. The magnitude of this sink is small and here we simply include it in the fossil fuel emission estimates as suggested in \citet[][p. 3277]{GCB2020short}.} The data series all start in $1959$ and end in $2016$, $2017$, $2018$, and $2019$, respectively (the data are collected with a one-year lag). We see that especially $S_t^\textrm{OCN}$ and $S_t^\textrm{LND}$, but also $E_t^\textrm{FF}$ and $E_t^\textrm{LUC}$, are revised when the new version of the Global Carbon Budget data set is published by the GCP. The bottom row of Figure \ref{fig:gcb} plots the budget imbalance \eqref{eq:bi} from each of the four data sets; again the revisions are clearly seen. The following section presents the results of statistical analyses of the budget imbalance data series shown in Figure \ref{fig:gcb}. It also investigates what is a satisfactory statistical model for the budget imbalance.

\vspace{0.75cm}
\centerline{[FIGURE \ref{fig:gcb} ABOUT HERE]}
\vspace{0.75cm}

\subsection{Statistical analysis of the budget imbalance}\label{sec:bi}
The upper panel of Table \ref{tab:BI} presents descriptive statistics regarding the four budget imbalance data series.  We find that the means of the time series are not significantly different from zero, indicating that the carbon budget has been balanced on average (Electronic Supplementary Material, Section 3). Further, the Durbin-Watson ($DW$) and Ljung-Box ($Q$) test statistics indicate that the budget imbalance contains (positive) serial autocorrelation. (The caption of Table \ref{tab:BI} contains additional information regarding the $DW$ and $Q$ statistics.) This, together with the visual impression of the bottom plot of Figure  \ref{fig:gcb}, provides a first indication of the budget imbalance being well-described by a zero-mean stationary process with some positive correlation structure. In the Electronic Supplementary Material, we report the results from conducting two different statistical tests of stationarity,  the ``KPSS'' test of \cite{KPSS1992} and the augmented Dickey-Fuller (``ADF'') test \citep[][]{DF1979}. These tests corroborate that the budget imbalance constitute a stationary process. This provides further evidence that the budget imbalance data are historically well-described by a stationary process.

Inspecting the empirical autocorrelation and partial autocorrelation functions of the budget imbalance data (not shown here for brevity, but given in the Electronic Supplementary Material), provides evidence that an autoregressive process of order one (AR(1)) is an adequate statistical model for $B_t^\textrm{IM}$ for all three data sets under study. Likewise, the Bayesian Information Criterion \citep[][]{BIC1978} selects an AR$(1)$ model from the class of autoregressive moving average (ARMA) models \citep[we refer to][for a textbook treatment of ARMA models]{Hamilton1994}. 

After fitting an AR($1$) model to the data, we subject the standardized residuals of this fit to the same analysis as conducted on the budget imbalance data in the beginning of this section. The results are shown in the bottom panel of Table \ref{tab:BI}. Again, the KPSS and ADF tests point towards these series being stationary for the four data sets (Electronic Supplementary Material). The residuals appear to have zero mean, unit standard deviation, and although their kurtosis is slightly less than $3$, the Jarque-Bera test statistic ($N$) implies that we cannot reject the null of Gaussianity (Table \ref{tab:BI}). A similar conclusion is reached using two other tests of Gaussianity, the Kolmogorov-Smirnov test (KS) and the Anderson-Darling test (AD). Indeed, in no case can we reject the null of Gaussianity at a $5\%$ level. In the Electronic Supplementary Material we provide QQ-plots of the data, which also corroborate these findings. Further, as evidenced by the $DW$ and $Q$ statistics, after fitting this model, there is practically no autocorrelation left in the residuals (Table \ref{tab:BI}). Thus, the statistical analysis is consistent with the standardized AR(1) residuals being an iid $N(0,1)$ sequence, providing further evidence of the adequateness of the AR(1) model for the budget imbalance data.

Summing up, the diagnostics confirm that the AR(1) model is a good model for the historical budget imbalance data studied here. The estimates (obtained by an ordinary least squares regression) of the autoregressive parameter $\hat{\phi}$ and for the standard deviation of the error terms $\hat{\sigma}$ do not vary much between the four data sets (Table \ref{tab:BI}). However, it is important to note that these parameters will not necessarily remain constant in future editions of the GCP data set. In particular, as discussed above, new modelling or accounting procedures might revise the budget imbalance data and in this way cause the statistical properties of the data, and thus the parameters describing them, to change. For instance, better estimates of the Earth system variables might result in a less autocorrelated and less variable budget imbalance, i.e. to a reduction in the autoregressive parameter $\phi$ and/or standard deviation parameter $\sigma$.

\vspace{0.75cm}

\centerline{[TABLE \ref{tab:BI} ABOUT HERE]}

\vspace{0.75cm}

\subsection{The budget imbalance when emissions are under-reported}\label{sec:mis}
Suppose that from some time point $\tau$, anthropogenic CO$_2$ emissions from fossil fuel consumption are misreported as the amount $E_t^{\textrm{FF},*}$, while the true amount emitted to the atmosphere is $E_t^\textrm{FF} \neq E_t^{\textrm{FF},*}$. Then, for $t \geq \tau$, the observed budget imbalance data become
\begin{linenomath}
\begin{align*}
B_t^{\textrm{IM},*} &= E_t^{\textrm{FF},*} + E_t^\textrm{LUC} - G_t^\textrm{ATM} - S_t^\textrm{OCN} - S_t^\textrm{LND} \\
	&= u_t + \xi_t,
\end{align*}
\end{linenomath}
where
\begin{linenomath}
\begin{align*}
u_t  = E_t^\textrm{FF} + E_t^\textrm{LUC} - G_t^\textrm{ATM} - S_t^\textrm{OCN} - S_t^\textrm{LND}, 
\end{align*}
\end{linenomath}
is the budget imbalance under the true (unobserved) emission path $E_t^\textrm{FF}$, while
\begin{linenomath}
\begin{align*}
\xi_t = E_t^{\textrm{FF},*}-E_t^\textrm{FF}, 
\end{align*}
\end{linenomath}
denotes the amount of misreporting of fossil fuel CO$_2$ emissions at time $t \geq \tau$.

In this case, the budget imbalance will take the form
\begin{linenomath}
\begin{align}\label{eq:BI}
B_t^{\textrm{IM},*} = \begin{cases}
    u_t       & \quad  t < \tau, \\
    u_t + \xi_t  & \quad t \geq \tau.
  \end{cases}
\end{align}
\end{linenomath}
In the previous section, we saw that the available data on the budget imbalance indicate that it is historically well-described by a mean-zero stationary AR(1) process. Equation \eqref{eq:BI} shows that if at some (unknown) time $t= \tau$, emissions start to become misreported, the budget imbalance time series will undergo a \emph{structural break}: it will go from being a zero-mean stationary process $u_t$ to being the sum of this process and the term $\xi_t = E_t^{\textrm{FF},*}-E_t^\textrm{FF}$, the latter possibly having non-zero mean. Indeed, since we expect any misreportings of CO$_2$ emissions to be under-reportings, that is $E_t^{\textrm{FF},*}<E_t^\textrm{FF}$, it will likely be the case that $\xi_t < 0$ for $t \geq \tau$, i.e. the structural break term will be \emph{negative}.

This motivates a monitoring procedure, where we perform sequential (yearly) tests of the null hypothesis that the budget imbalance data will continue to have zero mean, against the alternative hypothesis that a term with negative mean has been introduced into the observations. A rejection of this null hypothesis provides evidence that the data on the carbon fluxes $E_t^\textrm{FF}, E_t^\textrm{LUC}, G_t^\textrm{ATM}, S_t^\textrm{OCN}, S_t^\textrm{LND}$  are not compatible with the global  carbon budget equation \eqref{eq:be0}. In particular, as shown in this section, this will be the case if emissions are systematically under-reported.

\subsection{Practical data considerations}\label{sec:difficult}
As the discussion in the previous section shows, an implicit assumption of this paper is that the data on the CO$_2$ flux variables, $E_t^{\textrm{LUC}}, G_t^{\textrm{ATM}}, S_t^{\textrm{OCN}}, S_t^{\textrm{LND}}$, are collected in a way that is reasonably independent on data of reported fossil fuel emissions, $E_t^{\textrm{FF},*}$. Importantly, the data on the sinks, $S_t^{\textrm{OCN}}$ and $S_t^{\textrm{LND}}$, must not mechanically be ``subsuming'' any systematic misreporting in the CO$_2$ emissions data. Indeed, if this was the case, it would be possible to balance the global carbon budget \eqref{eq:be0} even in the presence of systematic misreportings of fossil fuel emissions, by adjusting $S_t^{\textrm{OCN}}$ and $S_t^{\textrm{LND}}$ accordingly. Consequently, in this case, it would not be possible to detect misreporting of fossil fuel emissions using the methods proposed in this paper. 

For these reasons, it is important that data on the CO$_2$ fluxes, especially $S_t^{\textrm{OCN}}$ and $S_t^{\textrm{LND}}$, are estimated using models which do not automatically balance the carbon budget equation \eqref{eq:be0} using data on reported emissions, $E_t^{\textrm{FF},*}$. In fact, such an approach, where the CO$_2$ flux of the sinks is estimated by accounting for $E_t^{\textrm{FF},*}$,  is often used in atmospheric inversion methods \citep[][p. 3288]{GCB2020short}. Therefore, we do not consider data obtained by such atmospheric inversion methods in this paper. Instead we consider the $S_t^{\textrm{OCN}}$ data obtained from a mixture of observation-based methods and estimates from global ocean biochemistry models, while the $S_t^{\textrm{LND}}$ data are from dynamic global vegetation models \citep[we refer to][pp. 3285--3288, for details]{GCB2020short}.

\section{Designing a sequential testing procedure for monitoring the budget imbalance}\label{sec:design}
Let $n \geq  1$ and denote by $\{y_t^{n}\}_{t=1}^n$ observations of a triangular array, given by
\begin{linenomath}
\begin{align*}
y_t^{n} = \begin{cases}
    u_t^n       & \quad  t = 1, 2, \ldots, \tau-1, \\
    u_t^n + \xi_t^n  & \quad t = \tau, \tau+1, \ldots, 
  \end{cases}
\end{align*}
\end{linenomath}
where $u_t^n$ is a zero-mean stationary stochastic process, $\xi_t^n$ is a process with non-zero mean, and $\tau$ is an unknown date where a ``structural break'' occurs. Here, $n$ denotes the number of observations in a specific ``vintage'' of the observations $\{y_t^{n}\}_{t=1}^n$; for instance, in the context of the GCP data studied in the previous section, we would have $n = 58$, $n = 59$, $n=60$, and $n=61$ for the GCB2017, GCB2018, GCB2019, and GCB2020 data sets, respectively, see Table \ref{tab:BI}. 

For all $n$, we assume that $y_t^n$ is observed over an initial time period $t= 1, \ldots, K$, with $\tau > K$, such that $\{y_t^{n}\}_{t=1}^K= \{u_t^{n}\}_{t=1}^K$. The initial period is used to estimate the statistical properties of $u_t^n$ and then the monitoring algorithm is initiated from time $t = K+1$. At some later (unknown) time $\tau \geq K+1$, a structural break occurs and the process $\xi_t^n$ is introduced into the observations. If no structural break occurs, set $\tau = \infty$. In the application to monitoring the budget imbalance considered in this paper, $u_t^n$ denotes the budget imbalance if emissions were reported faithfully, while  $\xi_t^n$ will subsume any systematic mis-measurements of the carbon fluxes in the carbon budget equation \eqref{eq:be0}.

As a model of the budget imbalance data, we propose to use the very general ARMA framework, i.e. for $n\geq 1$,
\begin{linenomath}
\begin{align}\label{eq:arma}
u_t^n =  \sum_{i=1}^{p_n} \phi_{i,n} u_{t-i}^n + \sigma_n\sum_{j=1}^{q_n}  \psi_{i,n} \epsilon_{t-j}^n + \sigma_n \epsilon_t^n, \quad t = 1, 2, \ldots, n,
\end{align}
\end{linenomath}
where $p_n$ and $q_n$ are integers denoting the number of autoregressive and moving average terms, respectively,  $\phi_{i,n},  \psi_{i,n} \in \R$, $\sigma_n>0$, and $\epsilon_t^n \sim N(0,1)$  is an iid error sequence. The choice of $p_n$ and $q_n$ should be decided based on statistical considerations such as those presented in Section \ref{sec:bi}; we recommend using the BIC. Motivated by the findings of Section \ref{sec:bi}, for GCB2017, GCB2018, GCB2019, and GCB2020, we will illustrate our methods assuming that $u_t^n$ is given by a stationary AR(1) process ($p_n = 1$, $q_n=0$), i.e. for $n\geq 1$,
\begin{linenomath}
\begin{align}\label{eq:u}
u_t^n = \phi_n u_{t-1}^n + \sigma_n \epsilon_t^n, \quad t = 1, 2, \ldots, n,
\end{align}
\end{linenomath}
where $\phi_n \in (-1,1)$, $\sigma_n>0$, and $\epsilon_t^n \sim N(0,1)$ is an iid error sequence. 

Our goal is to design a  sequential testing scheme, which continually monitors the budget imbalance time series to check for the presence of $\xi_t^n$. To be precise, at each time period $n = K+1, K+2, \ldots, K+T$, we propose to conduct a statistical test for whether $n\geq \tau$, i.e. for whether a structural break has occurred at or before the current time $n$. In other words, we are concerned with the sequence of hypothesis tests
\begin{linenomath}
\begin{align}\label{eq:Hseq}
H_{0,n}:  \tau>n, \quad \textnormal{against} \quad H_{1,n}: \tau \leq n,
\end{align}
\end{linenomath}
for $n = K+1, K+2, \ldots, T+K$, where $T$ denotes the number of periods in which we plan to perform the statistical tests. Monitoring can be done over  a fixed time horizon ($T<\infty$) or over an indefinite time horizon ($T= \infty$).

A widely used monitoring scheme is to recursively calculate the cumulated sum (CUSUM) of the observations, e.g. $\sum_n y_n^n$, and reject the null when this sum exceeds some critical boundary \citep[e.g.][]{Hinkley1971,CSW1996}. However, in this case, the critical boundary will depend on the statistical properties of the data (typically the long-run variance of $\{u_t^n\}$), which here means the parameters $\phi_n$ and $\sigma_n$ in \eqref{eq:u}. Since we allow for these parameters to vary with each new observation (vintage), this ``standard'' approach becomes infeasible. Instead, we seek a ``pivotal'' approach, i.e. an approach where the critical boundary does not depend on the parameters of the underlying data generating process. As we explain presently, we propose to construct such a test using estimates of the residuals $ \epsilon_t^n$ of \eqref{eq:u} \citep[a testing approach relying on the residuals in this way is often called ``innovation-based'' in the literature on testing for structural break in time series, see, e.g,][]{AH2012}.

To formalize our approach, consider the CUSUM test statistic
\begin{linenomath}
\begin{align}\label{eq:Zn}
Z_n = \sum_{s=K+1}^n \hat \epsilon_{s}^{s}, \qquad n = K+1, K+2, \ldots, K+T,
\end{align}
\end{linenomath}
where $\hat \epsilon_s^s$ is the year-$s$ estimate of the final error term $\epsilon_s^s$ of the statistical model of the budget imbalance data $\{y_i^s\}_{i=1}^s$ from the model \eqref{eq:arma}. In the case of the AR(1) model \eqref{eq:u}, this is
\begin{linenomath}
\begin{align}\label{eq:epshat}
\hat \epsilon_s^s =  \frac{y_{s}^{s} - \hat \phi_s y_{s-1}^s}{ \hat \sigma_s}, \qquad s = K+1, K+2, \ldots, K+T,
\end{align}
\end{linenomath}
where $\hat \phi_n$ and $\hat \sigma_n$ are (consistent) estimates of $\phi_n$ and $\sigma_n$ obtained from the initial data $\{y_t^n \}_{t=1}^K$ which, by assumption, is equal to $\{u_t^n \}_{t=1}^K$. Consequently, when the model for $u_t^n$ is properly specified, we will have $(\hat \phi_n, \hat \sigma_n) \stackrel{P}{\rightarrow} (\phi_n,\sigma_n)$ as $K \rightarrow \infty$.\footnote{We write ``$\stackrel{P}{\rightarrow}$'' for convergence in probability.} This implies that, for $n < \tau$, it also holds that $\hat \epsilon_n^n \stackrel{P}{\rightarrow} \epsilon_n^n$ as  $K \rightarrow \infty$. The upshot is that when $K$ is large and $n< \tau$, i.e. no structural break has occurred, the test statistic $Z_n$ in \eqref{eq:Zn} will behave approximately as a sum of $n-K$ independent $N(0,1)$ variables. 

Conversely, when $ n \geq \tau$, the process $\xi_t^n$ will be introduced into the observations and we can no longer expect that   $\hat \epsilon_n^n$, calculated using \eqref{eq:epshat}, will provide a good approximation to $\epsilon_n^n$ no matter how large $K$ is. To analyze this case further, consider the AR(1) model \eqref{eq:u} and note that for $s>\tau$ we have $y_s^s = u_s^s + \xi_s^s$. Thus, for large $K$ and using \eqref{eq:epshat}, it holds that
\begin{linenomath}
\begin{align*}
\hat \epsilon_s^s &=  \frac{y_{s}^{s} - \hat \phi_s y_{s-1}^s}{ \hat \sigma_s} = \frac{u_{s}^{s} - \hat \phi_s u_{s-1}^s}{ \hat \sigma_s} + \frac{\xi_{s}^s - \hat \phi_s \xi^s_{s-1}}{ \hat \sigma_s} \approx \epsilon_s^s + \overline{\xi_s^{s}},
\end{align*}
\end{linenomath}
where 
\begin{linenomath}
\begin{align*}
\overline{\xi_s^{s}} = \frac{\xi_{s} - \hat \phi_s \xi^s_{s-1}}{ \hat \sigma_s}.
\end{align*}
\end{linenomath}
Supposing that $\hat \phi_n \in (-1,1)$ and $0 > \xi^s_{s-1} \geq \xi^s_s$, the above implies $\overline{\xi_s^{s}} < 0$, i.e. a negative process has been introduced into the test statistic \eqref{eq:Zn}. From this discussion, the influence of the statistical parameters $\phi_n$ and $\sigma_n$ on the test under the alternative, i.e. when there is systematic under-reporting of CO$_2$ emissions,  also becomes clear. Indeed, the magnitude of the structural break process $\overline{\xi_s^{s}}$ will increase when $\phi_s$ and/or $\sigma_s$ decreases. The upshot is that better measurements of the Earth system variables, in the sense that the autocorrelation and variability of the budget imbalance decreases, will magnify the influence of $\overline{\xi_s^{s}}$ on the test statistic \eqref{eq:Zn}. In other words, we would expect that decreasing $\phi_n$ and/or $\sigma_n$ will result in better properties of the test under the alternative, i.e. when the null is false. In the Electronic Supplementary Material, we confirm this conjecture using simulations and find that the standard deviation parameter $\sigma_n$ is especially important: when $\sigma_n$ is lowered, the expected detection time of under-reportings of CO$_2$ emissions declines noticeably.

Since we are proposing to conduct many (sequential) hypothesis tests, cf. Equation \eqref{eq:Hseq}, the testing procedure must to be designed to avoid ``multiple testing'' problems, i.e. the fact that, if a fixed critical value is used, the probability of rejecting the null will tend to one, as the number of sequential tests grows \citep[e.g.][]{CSW1996}. That is, to make sure that the overall test has the desired size, the critical value used at each time point needs to be appropriately chosen. A natural way to do this, is to let the critical values increase as a function of time in an appropriate way. To formalize this, we introduce the \emph{boundary function} or \emph{critical value function} 
\begin{linenomath}
\begin{align}\label{eq:Ct}
C_n^{\alpha} = c_{T, \alpha}\cdot  f(n-K), \qquad n = K+1, K+2, \ldots, K+T,
\end{align}
\end{linenomath}
where $f(t)$ is a non-negative function and the constant $c_{T, \alpha}>0$ depends on the length of the monitoring period, $T$, and the chosen significance level for the test of $H_0$, $\alpha$. The boundary function $f$ can be specified by the researcher, under the condition that if $T = \infty$, then $f(t)/\sqrt{t} \rightarrow \infty$ as $t\rightarrow \infty$. Below, we use $T<\infty$, and choose the following simple functional form
\begin{linenomath}
\begin{align*}
f(t) = \sqrt{t}, \qquad t = 1, 2, \ldots.
\end{align*}
\end{linenomath}
We experimented with a number of different boundary functions, but did not find large differences in the performance of the resulting testing procedure.

For $n = K+1, K+2, \ldots, K+ T$, we now propose to conduct a series of one-sided sequential tests, using the rule 
\begin{linenomath}
\begin{align*}
\textnormal{``reject $H_{0,n}$ in favor of $H_{1,n}$ if $Z_n  \leq -C_n^{\alpha}$''.}
\end{align*}
\end{linenomath}
To ensure that the overall test 
\begin{linenomath}
\begin{align}\label{eq:H0}
H_{0}:  \tau>T, \quad \textnormal{against} \quad H_{1}: \tau \leq T,
\end{align}
\end{linenomath}
implemented  by sequentially testing $H_{0,n}$ for $n = K+1, K+2, \ldots, K+T$, asymptotically (as $K \rightarrow \infty$)  has the correct overall size $\alpha$, it is necessary to choose $c_{T,\alpha}$ such that
\begin{linenomath}
\begin{align*}
\mathbb{P}\left( \sum_{s=1}^t \epsilon_s \leq - C_t^{\alpha} , \textnormal{  for at least one $t \in \{1,2,\ldots, T\}$} \right) = \alpha,
\end{align*}
\end{linenomath}
where $\epsilon_s \sim N(0,1)$ are iid. Note that we here have formulated the hypothesis test as a one-sided test, since we saw above that it is negative values of $Z_n$ that are relevant to the alternative hypothesis (i.e. under-reporting of CO$_2$ emissions).

When $T = \infty$, it is possible to obtain a closed-form expression for $c_{T,\alpha}$ for certain boundary functions $f$ \citep[see, e.g.,][for examples]{CSW1996}.  In practice, however, for given $\alpha$, $T$, and boundary function $f$, we suggest to approximate $c_{T,\alpha}$ using Monte Carlo simulation as follows. Choose $B \geq 1$. For $b = 1, 2, \ldots, B$, simulate $T$ $N(0,1)$ variables $\{\epsilon^{(b)}_s\}_{s=1}^T$ and form the scaled sums $g_{(b)}(t) = (\sum_{s=1}^t \epsilon^{(b)}_s)/f(t)$, $t = 1, 2, \ldots, T$, and record the maximum $m_b = \max_{t \in \{1,2,\ldots,T\}} g_{(b)}(t)$. The approximated value of $c_{T,\alpha}$ is the $1-\alpha$ quantile of $\{m_b\}_{b=1}^B$. In our simulations, we set $B = 100\ 000$. If an indefinite  monitoring horizon ($T= \infty$) is desired, approximate $c_{\infty,\alpha}$ by setting $T$ to a very large value in the Monte Carlo simulation procedure, e.g. $T=1\ 000$. 

\begin{figure}[t!]
\centering
\includegraphics[width=1.0\linewidth]{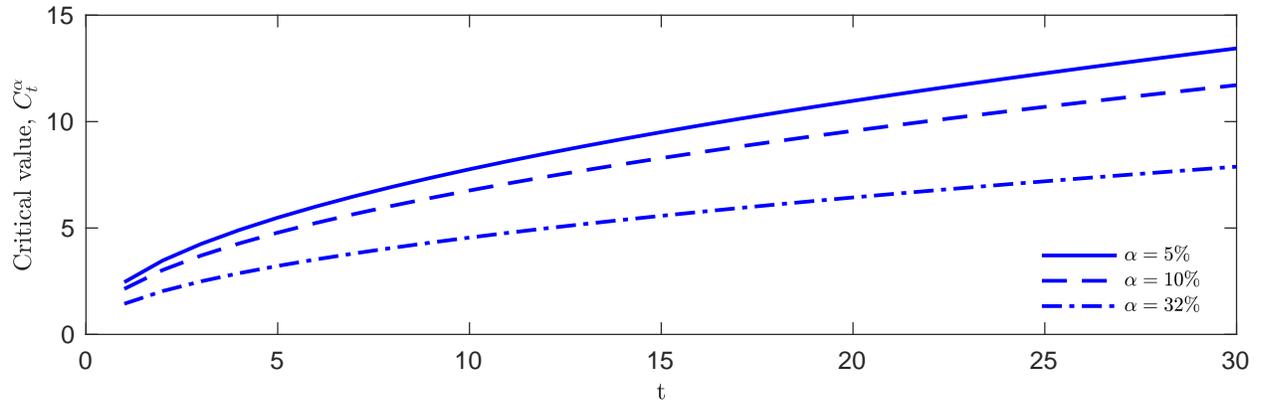} 
\caption{\emph{Critical value function $C_{K+t}^{\alpha}$ as a function of $t = 1, 2, \ldots, T$,  for $T = 30$ and significance level $\alpha = 5\%$ (solid line), $10\%$ (dashed line), $32\%$ (dashed and dotted line).}}
\label{fig:Ct}
\end{figure}

Figure \ref{fig:Ct} plots the critical value function \eqref{eq:Ct} for $T = 30$ and $\alpha = 5\%, 10\%, 32\%$.\footnote{Although, \emph{prima facie}, $\alpha = 32\%$ might seem like a high significance level to consider, a similar threshold is often used by the IPCC, where events happening with a probability lower than $33\%$ are termed ``unlikely'' \citep[][Table 1, p. 3]{IPCCuncertain}. Choosing such a high significance level will facilitate early detection of potential misreporting of CO$_2$ emissions; naturally, it comes with the caveat of a correspondingly high probability of making a Type I error.} In the context of this paper, a monitoring period of $T = 30$ years corresponds to the intention of using the proposed sequential testing procedure until around $2050$, if testing begins within the next few years. For instance, if the testing procedure is started when the data for $2020$ become available (e.g. in the forthcoming GCB2021 report), yearly testing could proceed until $2049$, which seems a rather fitting timeline in light of the Paris objectives (Section \ref{sec:future} contains further details on how to implement this).

\vspace{0.75cm}

\centerline{[FIGURE \ref{fig:Ct} ABOUT HERE]}
\vspace{0.75cm}

\subsection{Practical implementation of the sequential testing procedure}\label{sec:impl}
To implement the sequential testing procedure, it is necessary to first settle on a nominal significance level $\alpha \in (0,0.5]$ and a monitoring period $T\geq 1$, which could be indefinite ($T = \infty$). Using this, construct the boundary function $C_n^{\alpha}$ in \eqref{eq:Ct} using the approach outlined above. Now, starting from time $n = K+1$ until time $n = K+T$, the null hypotheses \eqref{eq:Hseq} are sequentially tested. In period $n$, this is done by first estimating the parametric model for the budget imbalance \eqref{eq:arma} using only the initial $K$ of the observations $\{y_t^n\}_{t=1}^K$. In the case of the AR(1) model \eqref{eq:u}, $\phi_n$ and $\sigma_n$ can be estimated by an OLS regression of $\{y_t^n\}_{t=2}^K$ on $\{y_t^n\}_{t=1}^{K-1}$. At this stage, it is recommended to verify the assumptions underlying the monitoring procedure, by testing whether the estimated residuals $\{ \hat \epsilon^n_t\}_{t=1}^K$ are approximately Gaussian, e.g. using the tests considered in Section \ref{sec:bi}. Now, using $\{ \hat \epsilon^s_s\}_{s=K+1}^n$, the test statistic $Z_n$ is calculated using \eqref{eq:Zn}. If, for some period $n$, the test statistic crosses the critical boundary, i.e. if $Z_n < - C_n^{\alpha}$, the null hypothesis $H_0$ in \eqref{eq:H0} is rejected and we stop. Rejection of $H_0$ in this way provides statistical evidence that the carbon flux data  $E_t^\textrm{FF}, E_t^\textrm{LUC}, G_t^\textrm{ATM}, S_t^\textrm{OCN}, S_t^\textrm{LND}$ are not compatible with the global carbon budget equation \eqref{eq:be0}.

\subsection{Simulation studies of the sequential testing procedure}\label{sec:sims} 
In the Electronic Supplementary Material, we investigate the finite sample properties of the proposed monitoring procedure when it is applied to simulated data. We set up the simulation study such as to mimic the setting of the Global Carbon Budget data studied in Section \ref{sec:gcb}. We briefly summarize the main findings here.

First, we find that the test is correctly sized, meaning when emissions are faithfully reported (i.e., $E_t^{\textrm{FF},*} = E_t^{\textrm{FF}}$), the probability of falsely rejecting the null (Type I error) is very close to the nominal size of the test, $\alpha$. That is, when $H_0$ given in \eqref{eq:H0} is true, the probability of rejecting the null is very close to $\alpha$ (Electronic Supplementary Material, Table 5).

\begin{figure}[t!]
\centering
\includegraphics[width=1.0\linewidth]{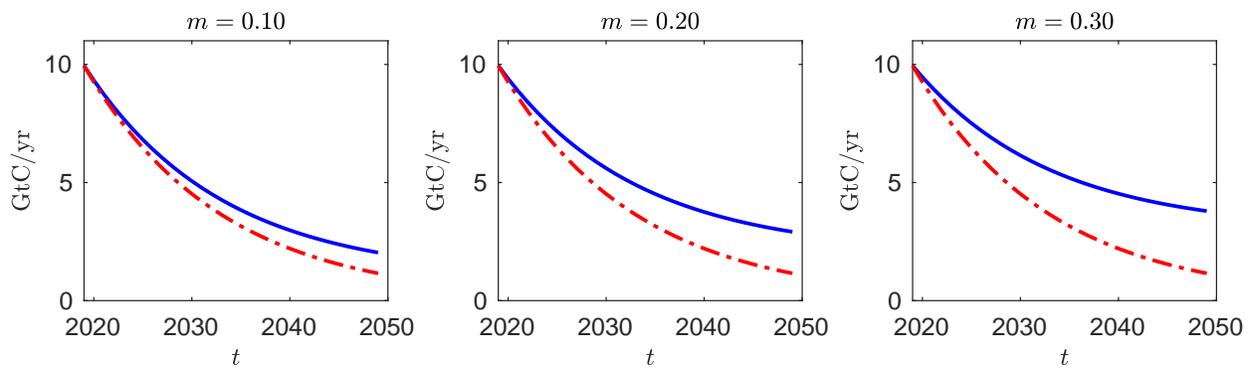} 
\caption{\emph{Example paths of actual emissions ($E_t^{\textrm{FF}}$; blue solid line) and reported emissions ($E_t^{\textrm{FF},*}$; red dashed line), for various values of misreporting parameter $m$, used in the simulation studies of power properties.}}
\label{fig:Epaths}
\end{figure}

We then investigate the power of the test, i.e. the ability of the test to reject $H_0$ when $H_0$ is false. These investigations require that we simulate the model under the alternative, i.e. under $H_1$. In the simulation studies, we assume that emissions are reported to decline according to the Paris objectives (we refer to the Electronic Supplementary Material for the precise details). For instance, we might assume that future emissions are reported to decline with a fixed fraction $g$ every year with respect to the current level of emissions, which would imply
\begin{linenomath}
\begin{align}\label{eq:F_report}
E_t^{\textrm{FF},*} = E_{2019}^{\textrm{FF}} (1-g)^{t-2019}, \quad t = 2020, 2021, \ldots.
\end{align}
\end{linenomath}
As argued in the Electronic Supplementary Material, we need to choose $g = 0.0692$ (corresponding to $6.92\%$ emissions abatement per year) to roughly adhere to the Paris objectives. Here, where $H_1$ is true, we have that future  actual emissions ($E_t^{\textrm{FF}}$) are greater than future reported emissions ($E_t^{\textrm{FF},*}$), i.e. $E_t^{\textrm{FF}} > E_t^{\textrm{FF},*}$ for $t  \geq 2020$. To model this, we introduce the \emph{misreporting parameter} $m\in [0,1]$, and set
\begin{linenomath}
\begin{align}\label{eq:F_actual}
E_t^{\textrm{FF}} =  (1-m)E_t^{\textrm{FF},*} + m E_{2019}^{\textrm{FF}}  , \quad t = 2020, 2021, \ldots.
\end{align}
\end{linenomath}
It is clear, that if $m = 0$, then $E_t^{\textrm{FF}} = E_t^{\textrm{FF},*}$ (no misreporting of CO$_2$ emissions), while if $m>0$, then $E_t^{\textrm{FF}} > E_t^{\textrm{FF},*}$ (under-reporting of CO$_2$ emissions).  Here, we can interpret the misreporting parameter $m$ to be the fraction of the emissions that are not abated, while a fraction $(1-m)$ of the emissions are abated according to the Paris objectives. This situation would obtain if  all countries report their fossil fuel emissions in a way which is consistent with the Paris objectives, but a number of countries, representing a fraction $m$ of emissions in $2019$, actually keep their emissions constantly equal to their $2019$ levels (i.e. they are under-reporting their emissions). Figure \ref{fig:Epaths} shows paths of $E_t^{\textrm{FF},*}$ and $E_t^{\textrm{FF}}$ for various values of the misreporting parameter $m$. It is clear that as $m$ grows, so does the amount of misreporting. Further, we see that the amount of misreporting is small in the beginning of the monitoring period and gradually grows as time progresses.

Using this setup in our simulation study, we find that the average detection time is quite large for the smaller amounts of under-reportings, i.e. for small values of $m$. Indeed, it is only for $m \geq 0.10$ that the power (probability of detecting under-reporting inside the monitoring period) becomes close to unity (Electronic Supplementary Material, Figures 7 and 8). The average detection time of the procedure will depend on the significance level $\alpha$ at which the test is conducted. A value of $m=0.20$ results in an average detection time between $7$ years ($\alpha = 32\%$) and $12$ years ($\alpha = 5\%$); for $m = 0.30$, the average detection time is between $5$ years ($\alpha = 32\%$)  and $10$ years ($\alpha = 5\%$); and for $m\geq 0.35$ the average detection time is smaller than $5$ years (for $\alpha = 32\%$). Further details are reported in the Electronic Supplementary Material.

We stress that these numbers should not be taken too literally. Indeed, the simulation study from which they arrive is highly stylized. Although conceptually simple and easy to understand, it is  unlikely that the path of reported and actual emissions will adhere precisely to \eqref{eq:F_report} and \eqref{eq:F_actual}. Instead, these simulation results should serve more as a ``proof-of-concept'', illustrating that the methods proposed in this paper can plausibly contribute to detecting potential future systematic under-reportings of CO$_2$ emissions. Likewise, the simulation study gives some qualitative indications that small amounts of under-reportings (here, $m \leq 0.10$) are difficult to detect; conversely, it appears that moderate-to-large under-reportings (here, $m \geq 0.20$) can potentially be detected quite quickly by the procedure proposed in this paper. Lastly, the simulations show that lowering the degree of autocorrelation (through the parameter $\phi_n$) or standard deviation (through the parameter $\sigma_n$) of the budget imbalance can result in substantial improvements of the properties of the test (Electronic Supplementary Material, Figures 7 and 8).

\section{Discussion and outlook}\label{sec:summary}
The sequential testing procedure proposed in Section \ref{sec:design} is very simple. Conceivably, the properties of the test can be improved by implementing a more advanced sequential testing procedure. Recent work in the literature on sequential ``on-line'' testing has devised numerous interesting procedures, such as ones where the critical boundaries are dynamically adjusted based on past tests \cite[e.g.,][]{RYWJ2017,RZWJ2018}, likelihood-based methods \cite[e.g.,][]{DG2020}, and methods that allow for open-ended monitoring periods \citep[e.g.,][]{GKD2021}, to name just a few examples. We note that, although such methods could potentially be useful in the present context, they would need to be adapted to the peculiar data structure encountered in our application, i.e. the fact that the data might be retroactively revised. Although interesting, we leave such adaptations of more advanced sequential testing algorithms, and their application to the present problem, for future research. 

It is important to keep in mind that  systematic under-reportings of CO$_2$ emissions are not the only possible reason for rejection of the null hypothesis proposed in this paper. Besides the inherent possibility of a false positive (Type I error), rejection of the null can also be caused by structural breaks in $E_t^\textrm{LUC}$ or the Earth system variables $G_t^\textrm{ATM}, S_t^\textrm{OCN}$, and $S_t^\textrm{LND}$. In particular, because the data series on land-use change emissions, $E_t^\textrm{LUC}$, and on the carbon sinks, $S_t^\textrm{OCN}$ and $S_t^\textrm{LND}$, are averages constructed from several advanced and complicated Earth system models, biases in these models could also result in rejection of the null. For these reasons, we recommend that following a rejection of the sequential null hypothesis, \emph{all} data series in the carbon budget equation \eqref{eq:be0} should be examined further. If the Earth system data series are deemed reliable, the statistical test provides evidence for CO$_2$ emissions being under-reported.

In a simulation study (Section \ref{sec:sims}; Electronic Supplementary Material), we sought to shed light on the empirical properties of the proposed monitoring procedure, with particular focus on how such a procedure might perform in practice for the case of verifying global CO$_2$ emissions in the future. The main take-aways from the simulation study were as follows:
\begin{enumerate}
\item
	For small under-reportings of CO$_2$ emissions, the power of the proposed monitoring device is low and the average detection time is high. This indicates that, if systematic under-reportings of CO$_2$ emissions are small, detection time can be long using the methods proposed in this paper. In particular, average detection time will likely (but not certainly) be too long to inform the $5$ yearly stocktakes of the Paris Agreement. 
\item
	For moderate-to-large under-reportings of CO$_2$ emissions, the proposed method has power close to unity and a short average detection time. That is, when under-reporting of CO$_2$ emissions is moderate-to-large, the proposed sequential testing procedure will detect this with high probability and in a timely enough manner that it can be used to inform the $5$ yearly stocktakes of the Paris Agreement.
\item
	The properties (power and detection time) of the proposed test improve dramatically when halving the standard deviation of the error term driving the budget imbalance (compare DGP3 with DGP1 in Figures 7 and 8 in the Electronic Supplementary Material). In other words, a more constrained and less volatile budget imbalance will make potential under-reportings much easier to detect.  This highlights the importance of continuing the effort to improve the precision of the data on Earth system variables, e.g. by refining the climate models used to obtain the data and/or by collecting more and higher quality observational data \citep[see also][for a similar conclusion]{Peters2017}. Note, however, that if such a sudden shift in data quality was to obtain, then it would be important to verify the underlying (stationarity) assumption on the budget imbalance. If this assumption is violated, the method should be altered accordingly (the key being that $\hat \epsilon_n^n$ is an approximately iid $N(0,1)$ sequence under the null).
\end{enumerate}

These observations indicate that the CO$_2$ monitoring procedure proposed in this paper cannot stand alone as a method of detecting potential future under-reportings of CO$_2$ emissions. However, they do indicate that it could possibly be a valuable tool in a larger portfolio of methods designed to verify global CO$_2$ emissions. For instance, \cite{CMZ2020} recently applied Benford's Law \citep{Benford1938}, a statistical accounting device which can be used to detect manipulation of reported numbers, to monitor emissions reduction claims of Clean Development Mechanism projects. The authors mention the possibility of applying Benford's Law to verifying reported global CO$_2$ emissions, i.e. for the same goal tackled in this paper.

Purely statistical methods, such as Benford's Law and the method proposed in this paper, are cheap and easy to apply. Although they do not guarantee fast detection of potential under-reporting, it seems prudent to implement these statistical methods along with other efforts for verifying reported CO$_2$ emissions, such as satellite-based monitoring of anthropogenic emissions, which is an area of current research and deployment  \citep[][]{Rajesh2016,Hakkarainen2016,Schwandner2017}. These ``indirect'' approaches should of course, as far as possible, be supplemented with direct investigations of the apparent reliability of individual reports of CO$_2$ emissions (or energy statistics) from individual nations. Such scrutinizing of national inventories are continuously being conducted, through independent technical expert reviews \citep[][]{UNFCCC1}, although currently only for developed (Annex I) countries  \citep[][]{UNFCCC2}.

\subsection{Monitoring future CO$_2$ emissions data}\label{sec:future}
Table \ref{tab:rule} presents the critical values, $C_n^{\alpha}$, for the test proposed in this paper when $T=30$. These are the critical boundaries which were used in the simulation experiment of Section \ref{sec:sims} and shown in Figure \ref{fig:Ct}. To monitor future CO$_2$ emissions, proceed as follows. Every year, when new data for $2020, 2021, \ldots$ arrive, calculate the budget imbalance using Equation \eqref{eq:bi} and update the test statistic $Z_n$ in Equation \eqref{eq:Zn} using the approach outlined in Section \ref{sec:impl} with $K = 61$, i.e. using the data up until $2019$ as the initial data series. Then compare $Z_n$ to the critical values given in Table \ref{tab:rule}: if, in some year, the test statistic is below the corresponding critical value, i.e. if $Z_n < -C_n^{\alpha}$, reject the null hypothesis \eqref{eq:H0} at the given significance level $\alpha$.

\vspace{0.75cm}

\centerline{[TABLE \ref{tab:rule} ABOUT HERE]}

\vspace{0.75cm}

These methods are easily implemented by any interested party using any reliable data set of the carbon fluxes in the carbon budget equation \eqref{eq:be0}, e.g. the data accompanying the yearly reports from the Global Carbon Project. The author will provide annual updates of this online.\footnote{\url{https://sites.google.com/site/mbennedsen/research/monitoring}.}

\section*{Acknowledgements}
The author would like to thank Eric Hillebrand, Siem Jan Koopman, three anonymous referees, and participants in the session on ``Climate change mitigation, impacts, and adaptation'' at the European Geoscience Union (EGU) General Assembly, Vienna, 2019, for many helpful comments and suggestions on the manuscript.

\section*{Declarations}

\noindent {\bf Funding:} Financial support from the Independent Research Fund Denmark for the project ``Econometric Modeling of Climate Change'' is acknowledged.

\noindent {\bf Conflicts of interest:} None.

\noindent {\bf Availability of data and material:}  The data used in this paper are freely available online; see \url{https://doi.org/10.18160/GCP-2020} for the GCB2020 version.

\noindent {\bf Code availability:} The MATLAB code used to produce the results of the paper is freely available at \url{https://sites.google.com/site/mbennedsen/research/monitoring}.

{\small 
\bibliographystyle{chicago}
\bibliography{gcb_references}

\begin{thebibliography}{}

\bibitem[\protect\citeauthoryear{Anderson and Darling}{Anderson and
  Darling}{1952}]{AD1952}
Anderson, T.~W. and D.~A. Darling (1952).
\newblock Asymptotic theory of certain "goodness of fit" criteria based on
  stochastic processes.
\newblock {\em Annals of Mathematical Statistics\/}~{\em 23\/}(2), 193--212.

\bibitem[\protect\citeauthoryear{Aue, H\"ormann, Horv\'ath, Hu\v{s}kov\'a, and
  Steinebach}{Aue et~al.}{2012}]{AHHHS2012}
Aue, A., S.~H\"ormann, L.~Horv\'ath, M.~Hu\v{s}kov\'a, and J.~G. Steinebach
  (2012).
\newblock Sequential testing for the stability of high-frequency portfolio
  betas.
\newblock {\em Econometric Theory\/}~{\em 28\/}(4), 804--837.

\bibitem[\protect\citeauthoryear{Aue and Horv\'ath}{Aue and
  Horv\'ath}{2012}]{AH2012}
Aue, A. and L.~Horv\'ath (2012).
\newblock Structural breaks in time series.
\newblock {\em Journal of Time Series Analysis\/}~{\em 34\/}(1), 1--16.

\bibitem[\protect\citeauthoryear{Benford}{Benford}{1938}]{Benford1938}
Benford, F. (1938).
\newblock The law of anomalous numbers.
\newblock {\em Proceedings of the American Philosophical Society\/}~{\em
  78\/}(4), 551--572.

\bibitem[\protect\citeauthoryear{Chu, Stinchcombe, and White}{Chu
  et~al.}{1996}]{CSW1996}
Chu, C.-S.~J., M.~Stinchcombe, and H.~White (1996).
\newblock Monitoring structural change.
\newblock {\em Econometrica\/}~{\em 64\/}(5), 1045--1065.

\bibitem[\protect\citeauthoryear{Cole, Maddison, and Zhang}{Cole
  et~al.}{2020}]{CMZ2020}
Cole, M.~A., D.~J. Maddison, and L.~Zhang (2020).
\newblock Testing the emission reduction claims of {CDM} projects using the
  {B}enford's {L}aw.
\newblock {\em Climatic Change\/}~{\em 160\/}(3), 407--426.

\bibitem[\protect\citeauthoryear{Dette and G{\"o}smann}{Dette and
  G{\"o}smann}{2020}]{DG2020}
Dette, H. and J.~G{\"o}smann (2020).
\newblock A likelihood ratio approach to sequential change point detection for
  a general class of parameters.
\newblock {\em Journal of the American Statistical Association\/}~{\em
  115\/}(531), 1361--1377.

\bibitem[\protect\citeauthoryear{Dickey and Fuller}{Dickey and
  Fuller}{1979}]{DF1979}
Dickey, D.~A. and W.~A. Fuller (1979).
\newblock Distribution of the estimators for autoregressive time series with a
  unit root.
\newblock {\em Journal of the American Statistical Association\/}~{\em
  74\/}(366a), 427--431.

\bibitem[\protect\citeauthoryear{Dlugokencky and Tans}{Dlugokencky and
  Tans}{2018}]{DT18}
Dlugokencky, E. and P.~Tans (2018).
\newblock Trends in atmospheric carbon dioxide.
\newblock National Oceanic \& Atmospheric Administration, Earth System Research
  Laboratory (NOAA/ESRL), available at:
  \url{http://www.esrl.noaa.gov/gmd/ccgg/trends/global.html}.

\bibitem[\protect\citeauthoryear{Duflo, Greenstone, Pande, and Ryan}{Duflo
  et~al.}{2013}]{DGPR2013}
Duflo, E., M.~Greenstone, R.~Pande, and N.~Ryan (2013).
\newblock Truth-telling by third-party auditors and the response of the
  polluting firms: Experimental evidence from {India}.
\newblock {\em Quarterly Journal of Economics\/}~{\em 128}, 1499--1545.

\bibitem[\protect\citeauthoryear{Durbin and Watson}{Durbin and
  Watson}{1971}]{Durbin1971}
Durbin, J. and G.~S. Watson (1971).
\newblock Testing for serial correlation in least squares regression.
\newblock {\em Biometrika\/}~{\em 58\/}(1), 1 -- 19.

\bibitem[\protect\citeauthoryear{Friedlingstein, Jones, and O'Sullivan~et
  al.}{Friedlingstein et~al.}{2019}]{GCB2019short}
Friedlingstein, P., M.~W. Jones, and M.~O'Sullivan~et al. (2019).
\newblock {Global Carbon Budget} 2019.
\newblock {\em Earth System Science Data\/}~{\em 11\/}(4), 1783--1838.

\bibitem[\protect\citeauthoryear{Friedlingstein, O'Sullivan, and Jones~et
  al.}{Friedlingstein et~al.}{2020}]{GCB2020short}
Friedlingstein, P., M.~O'Sullivan, and M.~W. Jones~et al. (2020).
\newblock {Global Carbon Budget} 2020.
\newblock {\em Earth System Science Data\/}~{\em 12\/}(4), 3269--3340.

\bibitem[\protect\citeauthoryear{Gasser, Crepin, Quilcaille, Houghton, Ciais,
  and Obersteiner}{Gasser et~al.}{2020}]{LUC2020}
Gasser, T., L.~Crepin, Y.~Quilcaille, R.~A. Houghton, P.~Ciais, and
  M.~Obersteiner (2020).
\newblock Historical {CO}$_{2}$ emissions from land use and land cover change
  and their uncertainty.
\newblock {\em Biogeosciences\/}~{\em 17\/}(15), 4075--4101.

\bibitem[\protect\citeauthoryear{Ghanem and Zhang}{Ghanem and
  Zhang}{2014}]{GZ2014}
Ghanem, D. and J.~Zhang (2014).
\newblock `{Effortless} {Perfection}': Do {Chinese} cities manipulate air
  pollution data?
\newblock {\em Journal of Environmental Economics and Management\/}~{\em
  68\/}(2), 203--225.

\bibitem[\protect\citeauthoryear{Gilfillan, Marland, Boden, and
  Andres}{Gilfillan et~al.}{2019}]{gilfillan2019}
Gilfillan, D., G.~Marland, T.~Boden, and R.~Andres (2019).
\newblock {Global, Regional, and National Fossil-Fuel CO2 Emissions}.
\newblock \url{https://energy.appstate.edu/CDIAC}.

\bibitem[\protect\citeauthoryear{G{\"o}smann, Kley, and Dette}{G{\"o}smann
  et~al.}{2021}]{GKD2021}
G{\"o}smann, J., T.~Kley, and H.~Dette (2021).
\newblock A new approach for open-end sequential change point monitoring.
\newblock {\em Journal of Time Series Analysis\/}~{\em 42\/}(1), 63--84.

\bibitem[\protect\citeauthoryear{Guan, Liu, Lindner, and Hubacek}{Guan
  et~al.}{2012}]{GLZH2012}
Guan, D., Z.~Liu, S.~Lindner, and K.~Hubacek (2012).
\newblock The gigatonne gap in {China} carbon dioxide inventories.
\newblock {\em Nature Climate Change\/}~{\em 2}, 672--675.

\bibitem[\protect\citeauthoryear{Hakkarainen, Ialongo, and
  Tamminen}{Hakkarainen et~al.}{2016}]{Hakkarainen2016}
Hakkarainen, J., I.~Ialongo, and J.~Tamminen (2016).
\newblock Direct space-based observations of anthropogenic {CO}2 emission areas
  from {OCO}-2.
\newblock {\em Geophysical Research Letters\/}~{\em 43\/}(21), 11,400--11,406.

\bibitem[\protect\citeauthoryear{Hamilton}{Hamilton}{1994}]{Hamilton1994}
Hamilton, J.~D. (1994).
\newblock {\em Time Series Analysis}.
\newblock Princeton University Press.

\bibitem[\protect\citeauthoryear{Hansis, Davis, and Pongratz}{Hansis
  et~al.}{2015}]{HDP15}
Hansis, E., S.~J. Davis, and J.~Pongratz (2015).
\newblock Relevance of methodological choices for accounting of land use change
  carbon fluxes.
\newblock {\em Global Biogeochemical Cycles\/}~{\em 29}, 1230 -- 1246.

\bibitem[\protect\citeauthoryear{Hinkley}{Hinkley}{1971}]{Hinkley1971}
Hinkley, D.~V. (1971).
\newblock Inference about the change-point from cumulative sum tests.
\newblock {\em Biometrika\/}~{\em 58\/}(3), 509--523.

\bibitem[\protect\citeauthoryear{Horv\'ath, Hu\v{s}kov\'a, Kokoszka, and
  Steinebach}{Horv\'ath et~al.}{2004}]{HHKS2004}
Horv\'ath, L., M.~Hu\v{s}kov\'a, P.~Kokoszka, and J.~Steinebach (2004).
\newblock Monitoring changes in linear models.
\newblock {\em Journal of Statistical Planning and Inference\/}~{\em 126\/}(1),
  225--251.

\bibitem[\protect\citeauthoryear{Houghton and Nassikas}{Houghton and
  Nassikas}{2017}]{HN17}
Houghton, R.~A. and A.~A. Nassikas (2017).
\newblock Global and regional fluxes of carbon from land use and land cover
  change 1850-2015.
\newblock {\em Global Biogeochemical Cycles\/}~{\em 31}, 456 -- 472.

\bibitem[\protect\citeauthoryear{IPCC}{IPCC}{2018}]{IPCC15}
IPCC (2018).
\newblock Special {Report} on {Global} {Warming} of 1.5$^{\circ}${C}.
\newblock Technical report, Intergovernmental Panel on Climate Change.

\bibitem[\protect\citeauthoryear{Janardanan, Maksyutov, Oda, Saito, Kaiser,
  Ganshin, Stohl, Matsunaga, Yoshida, and Yokota}{Janardanan
  et~al.}{2016}]{Rajesh2016}
Janardanan, R., S.~Maksyutov, T.~Oda, M.~Saito, J.~W. Kaiser, A.~Ganshin,
  A.~Stohl, T.~Matsunaga, Y.~Yoshida, and T.~Yokota (2016).
\newblock Comparing {GOSAT} observations of localized {CO}2 enhancements by
  large emitters with inventory-based estimates.
\newblock {\em Geophysical Research Letters\/}~{\em 43\/}(7), 3486--3493.

\bibitem[\protect\citeauthoryear{Jarque and Bera}{Jarque and
  Bera}{1987}]{JarqueBera1987}
Jarque, C.~M. and A.~K. Bera (1987).
\newblock A test for normality of observations and regression residuals.
\newblock {\em International Statistical Review\/}~{\em 2}, 163--172.

\bibitem[\protect\citeauthoryear{Korsbakken, Peters, and Andrew}{Korsbakken
  et~al.}{2016}]{KPA2016}
Korsbakken, J.~I., G.~P. Peters, and R.~M. Andrew (2016).
\newblock Uncertainties around reductions in {China's} coal use and {CO2}
  emissions.
\newblock {\em Nature Climate Change\/}~{\em 6}, 687--690.

\bibitem[\protect\citeauthoryear{Kwiatkowski, Phillips, Schmidt, and
  Shin}{Kwiatkowski et~al.}{1992}]{KPSS1992}
Kwiatkowski, D., P.~C. Phillips, P.~Schmidt, and Y.~Shin (1992).
\newblock Testing the null hypothesis of stationarity against the alternative
  of a unit root: How sure are we that economic time series have a unit root?
\newblock {\em Journal of Econometrics\/}~{\em 54\/}(1), 159--178.

\bibitem[\protect\citeauthoryear{Lai}{Lai}{1995}]{Lai1995}
Lai, T.~L. (1995).
\newblock Sequential changepoint detection in quality control and dynamical
  systems.
\newblock {\em Journal of the Royal Statistical Society: Series B\/}~{\em
  57\/}(4), 613--658.

\bibitem[\protect\citeauthoryear{Le~Qu\'er\'e, Andrew, and Friedlingstein~et
  al.}{Le~Qu\'er\'e et~al.}{2018a}]{GCB2017short}
Le~Qu\'er\'e, C., R.~M. Andrew, and P.~Friedlingstein~et al. (2018a).
\newblock {Global Carbon Budget} 2017.
\newblock {\em Earth System Science Data\/}~{\em 10\/}(1), 405 -- 448.

\bibitem[\protect\citeauthoryear{Le~Qu\'er\'e, Andrew, and Friedlingstein~et
  al.}{Le~Qu\'er\'e et~al.}{2018b}]{GCB2018short}
Le~Qu\'er\'e, C., R.~M. Andrew, and P.~Friedlingstein~et al. (2018b).
\newblock Global {Carbon} {Budget} 2018.
\newblock {\em Earth System Science Data\/}~{\em 10\/}(4), 2141--2194.

\bibitem[\protect\citeauthoryear{Ljung and Box}{Ljung and Box}{1978}]{LB1978}
Ljung, G.~M. and G.~E.~P. Box (1978).
\newblock On a measure of lack of fit in time series models.
\newblock {\em Biometrika\/}~{\em 65\/}(2), 297--303.

\bibitem[\protect\citeauthoryear{Luderer, Vrontisi, Bertram, Edelenbosch,
  Pietzcker, Rogelj, De~Boer, Drouet, Emmerling, Fricko, Fujimori, Havl{\'\i}k,
  Iyer, Keramidas, Kitous, Pehl, Krey, Riahi, Saveyn, Tavoni, Van~Vuuren, and
  Kriegler}{Luderer et~al.}{2018}]{Luderer2018}
Luderer, G., Z.~Vrontisi, C.~Bertram, O.~Y. Edelenbosch, R.~C. Pietzcker,
  J.~Rogelj, H.~S. De~Boer, L.~Drouet, J.~Emmerling, O.~Fricko, S.~Fujimori,
  P.~Havl{\'\i}k, G.~Iyer, K.~Keramidas, A.~Kitous, M.~Pehl, V.~Krey, K.~Riahi,
  B.~Saveyn, M.~Tavoni, D.~P. Van~Vuuren, and E.~Kriegler (2018).
\newblock Residual fossil {CO}2 emissions in 1.5--2$^{\circ}${C} pathways.
\newblock {\em Nature Climate Change\/}~{\em 8\/}(7), 626--633.

\bibitem[\protect\citeauthoryear{Marland and Rotty}{Marland and
  Rotty}{1984}]{MR1984}
Marland, G. and R.~M. Rotty (1984).
\newblock Carbon dioxide emissions from fossil fuels: a procedure for
  estimation and results for 1950--1982.
\newblock {\em Tellus B\/}~{\em 36B\/}(4), 232--261.

\bibitem[\protect\citeauthoryear{Massey}{Massey}{1951}]{Massey1951}
Massey, F.~J. (1951).
\newblock The {Kolmogorov-Smirnov} test for goodness of fit.
\newblock ~{\em 46\/}(253), 68--78.

\bibitem[\protect\citeauthoryear{Mastrandrea, Field, Stocker, Edenhofer, Ebi,
  Frame, Held, Kriegler, Mach, Matschoss, Plattner, Yohe, , and
  Zwiers}{Mastrandrea et~al.}{2010}]{IPCCuncertain}
Mastrandrea, M.~D., C.~B. Field, T.~F. Stocker, O.~Edenhofer, K.~L. Ebi, D.~J.
  Frame, H.~Held, E.~Kriegler, K.~J. Mach, P.~R. Matschoss, G.-K. Plattner,
  G.~W. Yohe, , and F.~W. Zwiers (2010).
\newblock Guidance note for lead authors of the {IPCC} {Fifth} {Assessment}
  {Report} on consistent treatment of uncertainties.
\newblock Technical report, {IPCC} Cross-Working Group Meeting on Consistent
  Treatment of Uncertainties.

\bibitem[\protect\citeauthoryear{Meinshausen, Jeffery, Guetschow, Robiou~du
  Pont, Rogelj, Schaeffer, H{\"o}hne, den Elzen, Oberth{\"u}r, and
  Meinshausen}{Meinshausen et~al.}{2015}]{Meinshausen2015}
Meinshausen, M., L.~Jeffery, J.~Guetschow, Y.~Robiou~du Pont, J.~Rogelj,
  M.~Schaeffer, N.~H{\"o}hne, M.~den Elzen, S.~Oberth{\"u}r, and N.~Meinshausen
  (2015).
\newblock National post-2020 greenhouse gas targets and diversity-aware
  leadership.
\newblock {\em Nature Climate Change\/}~{\em 5\/}(12), 1098--1106.

\bibitem[\protect\citeauthoryear{Millar, Fuglestvedt, Friedlingstein, Rogelj,
  Grubb, Matthews, Skeie, Forster, Frame, and Allen}{Millar
  et~al.}{2017}]{Millar2018}
Millar, R.~J., J.~S. Fuglestvedt, P.~Friedlingstein, J.~Rogelj, M.~J. Grubb,
  H.~D. Matthews, R.~B. Skeie, P.~M. Forster, D.~J. Frame, and M.~R. Allen
  (2017).
\newblock Emission budgets and pathways consistent with limiting warming to
  1.5$^{\circ}${C}.
\newblock {\em Nature Geoscience\/}~{\em 10}.

\bibitem[\protect\citeauthoryear{Nature}{Nature}{2018}]{Nature_Editorial2018}
Nature (2018, December 5).
\newblock Rules for a safe climate.
\newblock Editorial.

\bibitem[\protect\citeauthoryear{Page}{Page}{1954}]{page1954}
Page, E.~S. (1954).
\newblock Continuous inspection schemes.
\newblock {\em Biometrika\/}~{\em 41\/}(1), 100--115.

\bibitem[\protect\citeauthoryear{Peters, Le~Qu{\'e}r{\'e}, Andrew, Canadell,
  Friedlingstein, Ilyina, Jackson, Joos, Korsbakken, McKinley, Sitch, and
  Tans}{Peters et~al.}{2017}]{Peters2017}
Peters, G.~P., C.~Le~Qu{\'e}r{\'e}, R.~M. Andrew, J.~G. Canadell,
  P.~Friedlingstein, T.~Ilyina, R.~B. Jackson, F.~Joos, J.~I. Korsbakken, G.~A.
  McKinley, S.~Sitch, and P.~Tans (2017).
\newblock Towards real-time verification of {CO}$_2$ emissions.
\newblock {\em Nature Climate Change\/}~{\em 7\/}(12), 848--850.

\bibitem[\protect\citeauthoryear{Ramdas, F., Wainwright, and Jordan}{Ramdas
  et~al.}{2017}]{RYWJ2017}
Ramdas, A., Y.~F., M.~J. Wainwright, and M.~I. Jordan (2017).
\newblock Online control of the false discovery rate with decaying memory.
\newblock In {\em Advances in Neural Information Processing Systems}, pp.\
  5655--5664.

\bibitem[\protect\citeauthoryear{Ramdas, Zrnic, Wainwright, and Jordan}{Ramdas
  et~al.}{2018}]{RZWJ2018}
Ramdas, A., T.~Zrnic, M.~J. Wainwright, and M.~I. Jordan (2018).
\newblock {SAFFRON}: {A}n adaptive algorithm for online control of the false
  discovery rate.
\newblock In {\em Proceedings of the 35th International Conference on Machine
  Learning}, pp.\  4286--4294.

\bibitem[\protect\citeauthoryear{Reeves, Chen, Wang, Lund, and Lu}{Reeves
  et~al.}{2007}]{Reeves2007}
Reeves, J., J.~Chen, X.~L. Wang, R.~Lund, and Q.~Q. Lu (2007).
\newblock A review and comparison of changepoint detection techniques for
  climate data.
\newblock {\em Journal of Applied Meteorology and Climatology\/}~{\em 46\/}(6),
  900--915.

\bibitem[\protect\citeauthoryear{Sanderson, O'Neill, and Tebaldi}{Sanderson
  et~al.}{2016}]{SONBT2016}
Sanderson, B.~M., B.~C. O'Neill, and C.~Tebaldi (2016).
\newblock What would it take to achieve the {Paris} temperature targets?
\newblock {\em Geophysical Research Letters\/}~{\em 43\/}(13), 7133--7142.

\bibitem[\protect\citeauthoryear{Schwandner, Gunson, Miller, Carn, Eldering,
  Krings, Verhulst, Schimel, Nguyen, Crisp, O{\textquoteright}Dell, Osterman,
  Iraci, and Podolske}{Schwandner et~al.}{2017}]{Schwandner2017}
Schwandner, F.~M., M.~R. Gunson, C.~E. Miller, S.~A. Carn, A.~Eldering,
  T.~Krings, K.~R. Verhulst, D.~S. Schimel, H.~M. Nguyen, D.~Crisp, C.~W.
  O{\textquoteright}Dell, G.~B. Osterman, L.~T. Iraci, and J.~R. Podolske
  (2017).
\newblock Spaceborne detection of localized carbon dioxide sources.
\newblock {\em Science\/}~{\em 358\/}(6360).

\bibitem[\protect\citeauthoryear{Schwarz}{Schwarz}{1978}]{BIC1978}
Schwarz, G. (1978).
\newblock Estimating the dimension of a model.
\newblock {\em The Annals of Statistics\/}~{\em 6\/}(2), 461--464.

\bibitem[\protect\citeauthoryear{Tanaka and O'Neill}{Tanaka and
  O'Neill}{2018}]{TON2018}
Tanaka, K. and B.~C. O'Neill (2018).
\newblock The {Paris} {Agreement} zero-emissions goal is not always consistent
  with the 1.5$^{\circ}${C} and 2$^{\circ}${C} temperature targets.
\newblock {\em Nature Climate Change\/}~{\em 8\/}(4), 319--324.

\bibitem[\protect\citeauthoryear{Tokarska and Gillett}{Tokarska and
  Gillett}{2018}]{TG2018}
Tokarska, K.~B. and N.~P. Gillett (2018).
\newblock Cumulative carbon emissions budgets consistent with 1.5
  $\,^{\circ}${C} global warming.
\newblock {\em Nature Climate Change\/}~{\em 8\/}(4), 296--299.

\bibitem[\protect\citeauthoryear{{Transparency International}}{{Transparency
  International}}{2013}]{GlobalCorruption2013}
{Transparency International} (Ed.) (2013).
\newblock {\em Global Corruption Report: Climate Change}.
\newblock London: Routledge.

\bibitem[\protect\citeauthoryear{UN}{UN}{2017}]{ESyearbook}
UN (2017).
\newblock Energy statistics yearbook,
  \url{https://unstats.un.org/unsd/energystats/pubs/yearbook/}.
\newblock Technical report, United Nations Statistics Division.

\bibitem[\protect\citeauthoryear{UN}{UN}{2020}]{UNstats}
UN (2020).
\newblock Guidelines for the annual questionnaire on energy statistics,
  \url{https://unstats.un.org/unsd/energystats/questionnaire/documents/Energy-Questionnaire-Guidelines.pdf}.
\newblock Technical report, United Nations Statistics Division.

\bibitem[\protect\citeauthoryear{UNFCCC}{UNFCCC}{2015}]{FCCC2015}
UNFCCC (2015).
\newblock Adoption of the {Paris} {A}greement,
  \url{https://unfccc.int/resource/docs/2015/cop21/eng/l09r01.pdf}.
\newblock Technical report, United Nations Framework Convention on Climate
  Change.

\bibitem[\protect\citeauthoryear{UNFCCC}{UNFCCC}{2018a}]{Katowice2018}
UNFCCC (2018a).
\newblock Modalities, procedures and guidelines for the transparency framework
  for action and support referred to in {Article} 13 of the {Paris}
  {Agreement}.
\newblock Technical report, Ad Hoc Working Group on the {Paris Agreement},
  \url{https://unfccc.int/sites/default/files/resource/APA-SBSTA-SBI.2018.Informal.2.Add_.6_1.pdf}.

\bibitem[\protect\citeauthoryear{UNFCCC}{UNFCCC}{2018b}]{UNFCCC1}
UNFCCC (2018b).
\newblock Reporting and review under the {P}aris {A}greement,
  \url{https://unfccc.int/process-and-meetings/transparency-and-reporting/reporting-and-review-under-the-paris-agreement}.
\newblock Technical report, United Nations Framework Convention on Climate
  Change.

\bibitem[\protect\citeauthoryear{UNFCCC}{UNFCCC}{2018c}]{UNFCCC}
UNFCCC (2018c).
\newblock \url{https://di.unfccc.int/time_series}.

\bibitem[\protect\citeauthoryear{UNFCCC}{UNFCCC}{2019}]{UNFCCC2}
UNFCCC (2019).
\newblock Inventory review reports,
  \url{https://unfccc.int/process-and-meetings/transparency-and-reporting/reporting-and-review-under-the-convention/greenhouse-gas-inventories-annex-i-parties/inventory-review-reports-2019}.
\newblock Technical report, United Nations Framework Convention on Climate
  Change.

\bibitem[\protect\citeauthoryear{Unkel, Farrington, Garthwaite, Robertson, and
  Andrews}{Unkel et~al.}{2012}]{Unkel2012}
Unkel, S., C.~P. Farrington, P.~H. Garthwaite, C.~Robertson, and N.~Andrews
  (2012).
\newblock Statistical methods for the prospective detection of infectious
  disease outbreaks: a review.
\newblock {\em Journal of the Royal Statistical Society: Series A (Statistics
  in Society)\/}~{\em 175\/}(1), 49--82.

\bibitem[\protect\citeauthoryear{Zhang, Zhang, Qi, Huang, Karplus, and
  Zhang}{Zhang et~al.}{2019}]{Zhang2019}
Zhang, D., Q.~Zhang, S.~Qi, J.~Huang, V.~J. Karplus, and X.~Zhang (2019).
\newblock Integrity of firms' emissions reporting in {C}hina's early carbon
  markets.
\newblock {\em Nature Climate Change\/}~{\em 9\/}(2), 164--169.

\end{thebibliography}
}

\newpage

%
%

%
%
%
%
%
%
%

\begin{center}
\begin{table}[ht]
\caption{\emph{Descriptive statistics and diagnostics of the budget imbalance data from Equation \eqref{eq:bi}. $n$ is the number of observations and ``Mean'', ``Std'', ``Skew'', and ``Kurt'' are the empirical mean, standard deviation, skewness and kurtosis of the data. $\hat{\phi}$ and $\hat{\sigma}$ are the OLS regression estimates of the parameters in the AR(1) model $y_t = \phi y_{t-1} + \sigma \epsilon_t$, where $\epsilon_t \sim N(0,1)$. The data denoted ``AR(1) residuals'' are the estimated standardized error terms $\hat \epsilon_t :=( y_t - \hat \phi y_{t-1})/\hat \sigma$ from this regression. $N$ is the test-statistic from the Jarque-Bera test \citep[][]{JarqueBera1987}: the null hypothesis that the data comes from a Gaussian distribution can be rejected if $N$ is larger than the $5\%$ critical value of $5.99$. $KS$ is the Kolmogorov-Smirnov test statistic \citep[][]{Massey1951}: the null hypothesis that the data comes from a Gaussian distribution can be rejected if $KS$ is larger than the $5\%$ critical value of $0.18$. $AD$ is the Anderson-Darling test statistic \citep[][]{AD1952}: the null hypothesis that the data comes from a Gaussian distribution can be rejected if $AD$ is larger than the $5\%$ critical value of $0.74$. $DW$ is the Durbin-Watson test statistic \citep[][]{Durbin1971}: If $DW<2$ there is evidence of positive serial correlation in the data; if $DW>2$ there is evidence of negative serial correlation in the data; data without serial correlation will have $DW \approx 2$.  $Q(m)$ is the Ljung-Box Q test statistic \citep[][]{LB1978} for presence of autocorrelation calculated with $m$ lags. The $5\%$ critical value for the $Q(1)$-test is $3.84$; hence if $Q(1)>3.84$ then the null of no autocorrelation can be rejected at a $5\%$ level. The $5\%$ critical value for the Q(5) test is $11.07$.  All critical values assume a sample size of $58$ observations; all tests are implemented using built-un routines from the MATLAB programming language.}}
\label{tab:BI}
\footnotesize
\begin{tabularx}{.99\textwidth}{@{\extracolsep{\stretch{1}}}l|ccccccc|cccccc@{}} \hline
 & \multicolumn{7}{c}{Descriptive statistics} & \multicolumn{6}{c}{Diagnostics}  \\
 \cmidrule{2-8} \cmidrule{9-14} 
  &$n$& Mean & Std & Skew & Kurt & $\hat \phi$ & $\hat \sigma$  & $N$ & $KS$ & $AD$ & $DW$ & $Q(1)$ & $Q(5)$   \\ 
\emph{Budget imbalance}  \\
GCB2017   &      $58$  &    $0.00 $  &    $   0.86 $  &    $   0.02  $  &    $  2.44  $  &    $  0.44$    & $0.77$ &    $ 0.77   $  &  $0.07  $ & $ 0.18 $ &   $ 1.11  $  &    $ 11.90  $  &    $ 15.99    $     \\
GCB2018    &     $59 $  &   $0.14  $  &    $  0.80 $  &    $   0.04  $  &    $  2.41  $  &    $  0.43$  & $0.73$   &    $ 0.88  $  &    $ 0.12 $ & $ 0.40 $ &      $  1.11  $  &    $ 10.93  $  &    $ 14.87    $       \\
GCB2019     &   $ 60 $   &  $0.17 $  &    $   0.77  $  &    $ -0.07  $  &    $  2.81   $  &    $ 0.38$  & $0.72$   &    $ 0.14   $  &    $ 0.15 $ & $ 0.30 $ &     $ 1.21   $  &    $ 7.95  $  &    $ 10.36   $    \\ 
GCB2020     &   $ 61 $   &  $-0.01$  &    $   0.77  $  &    $ -0.20  $  &    $  3.40   $  &    $ 0.35$  & $0.72$   &    $ 0.80   $  &    $ 0.12 $ & $ 0.30 $ &     $ 1.29   $  &    $ 7.70 $  &    $ 9.61   $    \\[0.2cm]
\emph{AR$(1)$ residuals} \\
GCB2017   &    $57  $  &    $  0.01  $  &    $  1.00 $  &    $  -0.14 $  &    $   2.07  $  &    $ -0.01 $  &   $1.01$&  $   2.24   $  &   $ 0.08   $ &    $  0.49  $   &          $ 2.01 $  &    $   0.01 $  &    $   0.07   $         \\
GCB2018   &    $58 $  &    $  -0.09  $  &    $  1.00 $  &    $  -0.13 $  &    $   2.13 $  &    $  -0.03  $  &   $1.01$ & $  1.98  $  &    $ 0.10   $ &    $ 0.41   $   &          $  2.05  $  &    $  0.08 $  &    $   0.27 $        \\
GCB2019   &   $59  $  &    $ -0.13 $  &    $   0.99 $  &    $   0.05  $  &    $  2.28  $  &    $ -0.03   $  &  $1.01$ &  $ 1.28  $  &   $ 0.11   $ &    $ 0.24   $   &            $  2.06  $  &    $  0.14   $  &    $ 0.65  $        \\
GCB2020  &   $60  $  &    $ 0.20$  &    $  1.00 $  &    $   0.21  $  &    $  2.80  $  &    $ -0.02   $  &  $1.01$ &  $ 0.54 $  &    $ 0.07   $ &    $  0.35  $   &            $  2.03 $  &    $  0.03  $  &    $ 1.67  $        \\  \hline
 \end{tabularx}
\end{table}
\end{center}

\begin{center}
\begin{table}[ht]
\footnotesize
\begin{tabularx}{.99\textwidth}{@{\extracolsep{\stretch{1}}}lcccccccccc@{}} \hline
  &    $2020$ &  $2021$ & $2022$ & $2023$ & $2024$ & $2025$ & $2026$ & $2027$ & $2028$  & $2029$  \\ 
   \cmidrule{2-11} \\[-0.5cm]
$\alpha = 5\%$ &  $ 2.45  $  &  $   3.47 $  &  $    4.25  $  &  $   4.91  $  &  $   5.49  $  &  $   6.01  $  &  $   6.49  $  &  $   6.94  $  &  $   7.36  $  &  $   7.76$  \\ 
$\alpha = 10\%$ &     $2.14 $  &  $    3.02  $  &  $   3.70  $  &  $   4.28  $  &  $   4.78  $  &  $   5.24  $  &  $   5.66  $  &  $   6.05   $  &  $  6.42   $  &  $  6.76 $\\ 
$\alpha = 32\%$  &  $   1.44 $  &  $    2.03  $  &  $   2.49   $  &  $  2.88  $  &  $   3.22  $  &  $   3.52  $  &  $   3.81   $  &  $  4.07   $  &  $  4.32  $  &  $   4.55$ \\   \hline
 \end{tabularx}
\caption{\emph{Critical values $C_t^{\alpha}$ for the test of \eqref{eq:H0}. The calculations are made using $T= 30$, implying a monitoring period of $30$ years. Only the critical values for the first ten years are shown.}}\label{tab:rule}
\end{table}
\end{center}

\end{document}